\documentclass[11pt,xcolor=dvipsnames,fleqn]{article}
\DeclareMathAlphabet{\scr}{U}{rsfs}{m}{n}

\pdfoutput=1
\usepackage{scalerel,stackengine}
\usepackage[colorlinks,urlcolor=black,linkcolor = blue,citecolor = black,]{hyperref}
\usepackage{subfigure}
\usepackage{latexsym}
\usepackage{epsfig}
\usepackage[mathscr]{eucal}
\usepackage{amsfonts}
\usepackage{amscd}
\usepackage{cite}
\usepackage{array}
\usepackage{amssymb}
\usepackage{colordvi}
\usepackage[centertags]{amsmath}
\usepackage{enumerate}
\usepackage{graphicx}
\usepackage{booktabs}
\usepackage{theorem}
\usepackage[footnotesize]{caption}
\usepackage{soul}
\usepackage{mcite}
\usepackage{slashed}%slash notation for gamma matrices
\usepackage{braket}%for bra and ket notations
\usepackage{xcolor}
\usepackage{bbm}
\usepackage[utf8]{inputenc}
\usepackage{fancyvrb}% provides improved Verbatim command
\usepackage{framed}
\usepackage{xspace}
\usepackage{todonotes}
\usepackage{tikz-feynman}%Package to draw Feynman Diagrams
\usepackage{siunitx}%Allows for formatting units via \SI{number}{Unit}

\newcommand{\be}{\begin{equation}}
\newcommand{\ee}{\end{equation}}
\newcommand{\bea}{\begin{eqnarray}}
\newcommand{\eea}{\end{eqnarray}}

\newcommand{\beq}{\begin{eqnarray}}
\newcommand{\eeq}{\end{eqnarray}}
\newcommand{\bpmatrix}{\begin{pmatrix}}
\newcommand{\epmatrix}{\end{pmatrix}}
\newcommand{\ba}{\begin{array}}
\newcommand{\ea}{\end{array}}

\newcommand{\bc}{\begin{center}}
\newcommand{\ec}{\end{center}}

\newcommand{\ii}{\mathit{i}}

\clearpage{\pagestyle{empty}\cleardoublepage}

\newcommand{\cbrak}[1]{\left(#1\right)}

\newcommand{\X}{\chi}
\newcommand{\mX}{m_{\chi}}

\newcommand{\GG}{G_{\mu\nu}^aG^{a\mu\nu}}

\definecolor{blobColor}{RGB}{191,191,191}
\allowdisplaybreaks

\usepackage{cleveref}
\crefname{chapter}{Chapter}{Chapter}
\crefname{section}{Sec.}{Secs.}
\crefname{table}{Tab.}{Tabs.}
\crefname{figure}{Fig.}{Figs.}
\crefname{equation}{Eq.}{Eqs.}
\crefname{appendix}{Appendix\ }{Appendix\ }

\begin{document}
%\pdfoutput=1

\title{
    \vspace*{-3.7cm}
    \phantom{h} \hfill\mbox{\small KA-TP-11-2022}\\[-1.1cm]
    \vspace*{2.7cm}
    \textbf{Electroweak Corrections to \\ Dark Matter Direct Detection \\
        in the Dark Singlet Phase of the N2HDM\\[4mm]}}

\date{}
\author{
Seraina Glaus$^{1,2\,}$,
Margarete M\"{u}hlleitner$^{1\,}$\footnote{E-mail:
    \texttt{milada.muehlleitner@kit.edu}} ,
Jonas M\"{u}ller$^{1\,}$,
\\
Shruti Patel$^{1,2\,}$,
Rui Santos$^{3,4\,}$\footnote{E-mail:
    \texttt{rasantos@fc.ul.pt}}
\\[5mm]
{\small\it
$^1$Institute for Theoretical Physics, Karlsruhe Institute of Technology,} \\
{\small\it 76128 Karlsruhe, Germany}\\[3mm]
{\small\it
$^2$Institute for Nuclear Physics, Karlsruhe Institute of Technology,
76344 Karlsruhe, Germany}\\[3mm]
{\small\it $^3$Centro de F\'{\i}sica Te\'{o}rica e Computacional,
Faculdade de Ci\^{e}ncias,} \\
{\small \it    Universidade de Lisboa, Campo Grande, Edif\'{\i}cio C8
1749-016 Lisboa, Portugal} \\[3mm]
{\small\it
$^4$ISEL -
Instituto Superior de Engenharia de Lisboa,} \\
{\small \it   Instituto Polit\'ecnico de Lisboa
1959-007 Lisboa, Portugal} \\[3mm]
}

\maketitle

%%%%%%%%%%%%%%%%%%%%%%%%%%%%%%%%%%%%%%%%%%%%%%%%%%%%%%%%%%%%%%%%%%%%%%%%%%%%%%%%%%%%%%%%%%%%5
\begin{abstract}
Direct detection experiments are the only way to obtain indisputable
evidence of the existence of dark matter (DM) in the form of a particle. These experiments have been used to probe 
many extensions of the Standard Model (SM) that provide DM candidates. Experimental results like the latest ones from XENON1T
lead to severe constraints in the parameter space of many of the
proposed models. In a simple extension of the SM, the addition of a complex singlet to the SM content, 
one-loop corrections need to be taken into account because the tree-level cross section is proportional to the DM velocity, and therefore negligible. 
In this work we study the case of a DM particle with origin in a singlet but in a larger framework of an extension by an extra doublet together with the extra singlet providing the DM candidate.
We show that in the region of interest of the present and future direct detection experiments, electroweak corrections are quite stable with a $K$-factor very close to one.
\end{abstract}
\thispagestyle{empty}
\vfill
\newpage
\setcounter{page}{1}
%%%%%%%%%%%%%%%%%%%%%%%%%%%%%%%%%%%%%%%%%%%%%%%%%%%%%%%%%%%%%%%%%%%%%%%%%%%%%%%%%%%%%%%%%%%%
\section{Introduction}
%\cite{LUX-ZEPLIN:2018poe}
The only way to unmistakably identify a dark matter (DM) particle is in direct detection experiments.
In the mass region of the so-called Weekly Interacting Massive Particles (WIMPs) the latest and most restrictive
constraints were obtained by the XENON1T collaboration~\cite{Aprile:2017iyp,Aprile:2018dbl}. In this type of experiments, when DM interacts
with XENON it creates light and electric charge and theses signals provide information about the energy
and location of the initial collision. Since direct detection experiments play the major role in probing the WIMP
region it is important to understand in great detail the DM-nucleon cross sections in the different models.
There is a particularly interesting case, the one of the extension of the SM by a complex singlet that leads to
a tree-level DM-nucleon cross section proportional to the DM velocity
and therefore to a negligible rate~\cite{Gross:2017dan}.
The calculation of the electroweak corrections to DM-nucleon scattering in this model was performed in~\cite{Azevedo:2018exj, Ishiwata:2018sdi, Glaus:2020ihj}
and shown to be several orders of magnitude above the tree-level result.

In previous works we have also calculated the electroweak corrections~\cite{Glaus:2019itb, Glaus:2020ape} in a vector DM model~\cite{Azevedo:2018oxv}.
In this case the tree-level cross section is not negligible and electroweak corrections, in the region not excluded by XENON1T, are quite stable with
a $K$-factor close to 1 ($K=\sigma_{\text{NLO}}/\sigma_{\text{LO}}$ - the ratio of the next-to-leading order to the leading order cross section). In this letter
we discuss a scenario where  the DM candidate originates from a singlet but now within the larger
framework of the Dark Singlet Phase (DSP)~\cite{Engeln:2020fld, Engeln:2018ywp} of the next to 2-Higgs Doublet Model (N2HDM). 
The first point to note is that in this case there is no tree-level cancelation. Hence, the leading order cross section is not negligible.
%After discussing the cases of simple extensions like adding just one complex singlet to the SM or to have a new gauge boson from a dark $U(1)$ symmetry
%as a DM candidate we wanted to close this series of works by considering a model with a much larger parameter space. 
The main question we would like to answer is if the corrections are still stable and not too large when the parameter space of the visible sector
is enlarged which is the case of the DSP os the N2HDM. The DM candidate is singlet-like but the visible sector is now
a $Z_2$ symmetric 2HDM, with a new set of parameters and extra contributions to the electroweak corrections. As we have
discussed in great detail all the steps of the calculations  in our previous works~ \cite{Glaus:2019itb, Glaus:2020ape,  Glaus:2020ihj}  
 and also because there are no major changes in the methodology we will whenever possible refer the reader to those works
and will just focus on the differences for the model under study.

The outline of the letter is as follows. In section \ref{sec:model}, we will introduce the DSP of the N2HDM together with our
notation. Section~\ref{sec:ren} contains a brief description of the renormalization procedure used in this work. 
In section~\ref{sec:cxsi} we calculate the electroweak corrections to the spin-independent direct detection cross section. In section \ref{sec:num}, the results
are presented and discussed. Finally, we present our conclusions in section~\ref{sec:conc}.

\section{The Dark Singlet Phase of the N2HDM \label{sec:model}}
The model considered in this work is the DSP of the N2HDM
\cite{Chen:2013jvg,Muhlleitner:2016mzt,Muhlleitner:2017dkd}. The
Higgs sector of the SM is extended by one complex $SU(2)_L$ doublet
with hypercharge $+1$, and one real $SU(2)_L$ singlet with hypercharge $0$. We focus on a particular phase of the four possible dark phases of the N2HDM, the DSP, where
the singlet field has a vanishing vacuum expectation value (VEV) and
does not couple to the SM fields, making it a DM candidate. A detailed
discussion of the different dark matter 
phases of the N2HDM can be found in~\cite{Engeln:2020fld,Ferreira:2019iqb}. The Yukawa version of the model is type I meaning that that all quarks and leptons couple to only one of the doublets.
The Higgs potential is simplified by requiring invariance under the two $\mathbb{Z}_2$ symmetries,
\begin{align}
\mathbb{Z}^{(1)}_{2}&:\quad \Phi_{1}\rightarrow -\Phi_{1},\quad \Phi_{2}\rightarrow \Phi_{2},\quad \Phi_{S}\rightarrow \Phi_{S}\,,\label{eq:Z2_1}\\
\mathbb{Z}^{(2)}_{2}&:\quad \Phi_{1}\rightarrow \Phi_{1},\quad \Phi_{2}\rightarrow \Phi_{2},\quad \Phi_{S}\rightarrow -\Phi_{S}\label{eq:Z2_2} \, ,
\end{align}
which allows us to write the most general CP-conserving and renormalizable scalar potential invariant under these $\mathbb{Z}_2$ symmetries as
\begin{align}
V_{\text{scalar}} =&\enspace m_{11}^{2} \Phi_{1}^{\dagger} \Phi_{1} + m_{22}^{2} \Phi_{2}^{\dagger} \Phi_{2} + \dfrac{\lambda_{1}}{2} \left(\Phi_{1}^{\dagger} \Phi_{1}\right)^{2}
+ \dfrac{\lambda_{2}}{2} \left(\Phi_{2}^{\dagger} \Phi_{2}\right)^{2} \notag \\
&+\enspace \lambda_{3} \Phi_{1}^{\dagger} \Phi_{1} \Phi_{2}^{\dagger} \Phi_{2} + \lambda_{4} \Phi_{1}^{\dagger} \Phi_{2} \Phi_{2}^{\dagger} \Phi_{1}
+ \dfrac{\lambda_{5}}{2} \left[\left(\Phi_{1}^{\dagger} \Phi_{2}\right)^{2} + \text{h.c.}\right]\label{eq:scalpot}\\
&+\enspace \dfrac{1}{2} m_{s}^{2}\Phi_{S}^{2} + \dfrac{\lambda_{6}}{8} \Phi_{S}^{4} + \dfrac{\lambda_{7}}{2} \Phi_{1}^{\dagger} \Phi_{1} \Phi_{S}^{2} + \dfrac{\lambda_{8}}{2} \Phi_{2}^{\dagger} \Phi_{2} \Phi_{S}^{2}\,,\notag
\label{eq:Vscalar}
\end{align} 
with three real  mass dimension parameters $m_{11}^2, m_{22}^2, m_{s}^2$ and eight real dimensionless parameters $\lambda_{1} \cdots \lambda_{8}$. 
The symmetry $\mathbb{Z}^{(1)}_{2}$ is spontaneously broken by the doublet VEV. In the DSP, both doublets acquire VEVs but the singlet VEV vanishes 
which leaves the symmetry  $\mathbb{Z}^{(2)}_{2}$ unbroken. After electroweak symmetry breaking (EWSB) the doublet and singlet fields can be parametrized in terms of the VEVs $v_1$ and $v_2$, and component fields as
\be
\Phi_1 = \bpmatrix \phi_1^+ \\ \frac{1}{\sqrt{2}} (v_1 + \rho_1 + i\eta_1) \epmatrix \,, \qquad \Phi_2 = \bpmatrix \phi_2^+ \\ \frac{1}{\sqrt{2}} (v_2 + \rho_2 + i\eta_2) \epmatrix \,, \qquad
\Phi_s = \rho_s\,,
\label{eq:higgs-fields}
\ee
where $\phi_1^+, \phi_2^+$ are  complex charged fields, $\rho_1, \rho_2$, $\rho_s$ $\eta_1$ and $\eta_2$ are neutral fields. After EWSB, the CP-even fields $\rho_1$ and $\rho_2$ mix and give rise to the CP-even mass eigenstates $h_1$ and $h_2$ defined such that $m_{h_1} \leq m_{h_2}$, and either $h_1$ or $h_2$ can be identified with the 125~GeV SM Higgs boson. Similarly $\eta_1$ and $\eta_2$ mix to give a pseudoscalar mass eigenstate $A$ and the neutral Goldstone boson $G^0$. Finally, $\phi_1^+, \phi_2^+$ mix to give a charged Higgs $H^+$ and the charged Golstone boson $G^+$. The singlet field $\rho_s$ does not mix with any of the doublet fields, nor does it couple to any SM particles. Moreover, the unbroken $\mathbb{Z}^{(2)}_{2}$ symmetry gives rise to a dark parity, such that $\rho_s \equiv \chi$ with mass $m_\chi$ emerges as a DM candidate in the model. 

The mass eigenstates can be expressed in terms of the gauge eigenstates via rotation matrices as follows,
\begin{align}
\qquad \bpmatrix h_1 \\ h_2 \\ \chi \epmatrix = R_{\alpha} \bpmatrix \rho_1 \\ \rho_2 \\ \rho_s \epmatrix \,\,,\qquad \bpmatrix G^0 \\ A \epmatrix = U_\beta \bpmatrix \eta_1 \\ \eta_2 \epmatrix\,\,, \qquad \bpmatrix G^\pm \\ H^\pm \epmatrix = U_\beta \bpmatrix \phi_1^\pm \\ \phi_2^\pm \epmatrix\,,
\label{eq:mass-eigen}
\end{align}
where the rotation matrices are parametrized as\footnote{Note the different parametrization of $R_\alpha$ compared to \cite{Engeln:2020fld, Engeln:2018ywp}.}
\be
 R_\alpha = \bpmatrix \cos \alpha && \sin \alpha && 0 \\
					 -\sin \alpha && \cos \alpha && 0 \\
					  0           &&   0         && 1  \epmatrix \,\,, \qquad
U_\beta = \bpmatrix \cos \beta && \sin \beta \\
-\sin \beta && \cos \beta   \epmatrix\,.
\label{eq:rot-matrix}
\ee
The VEVs of the two doublets are related to the SM VEV ($v \approx 246.22$ GeV) as 
\begin{align}
v_1 = v \cos \beta \,\,,\qquad v_2 = v \sin \beta\,\,,
\end{align}
with $v = 2m_W/g$, where $m_W$ is the mass of the $W^\pm$ boson.

Further details on the minimization conditions can be found in~\cite{Engeln:2020fld, Engeln:2018ywp}. The final set of independent input parameters chosen is  
\begin{align}
v\,,\alpha\,,\tan \beta\,,m_{\chi} \,,m_{h_1}\,, m_{h_2}\,, m_A\,, m_{H^\pm}\,,\lambda_{6}\,,\lambda_{7}\,,\lambda_{8}\,.
\label{eq:inputs}
\end{align}
%\textcolor{blue}{Should explicit expressions for parameter transformations be added here?}

%%%%%%%%%%%%%%%%%%%%%%%%%%%%%%%%%%%%%%%%%%%%%%%%%%%%%%%%%%%%%%%%%%%%%%%%%%%%%%%%%%%%%%%%%%%%%%%%

\section{Renormalization of the Model \label{sec:ren}}

In the following, we present the renormalization of the N2HDM DSP in order to calculate the electroweak (EW) corrections to the scattering process of the scalar DM particle $\chi$
with a nucleon. We follow the prescription presented in~\cite{Krause:2017mal} by adapting the renormalization of the unbroken phase of the N2HDM to our scenario. This is done by taking the limit $v_s =  \alpha_2 = \alpha_3 = 0$ 
and by additionally renormalizing the parameters $\lambda_{7}$ and $\lambda_{8}$, which are the ones that enter our calculation.  

The bare input parameters $p_0$ defined in \cref{eq:inputs} are expressed in terms of the renormalized parameters $p$ and their respective counterterms $\delta p$ as 
\be
p_0  = p + \delta p\,,
\ee
whereas bare fields $\phi_0$ are expressed in terms of the renormalized fields $\phi$ and the wave-function renormalization factors (WFRs) $Z_{\phi}$ as
\be
\phi_0 = \sqrt{Z_{\phi}} \phi\,
\ee
where $Z_{\phi}$ is a matrix in case the fields mix at one-loop. The renormalization conditions define the finite parts of the counterterms. In this work, we will use the on-shell (OS) renormalization scheme to fix the 
renormalization constants for the masses and fields. The tadpoles are treated in the Fleischer and Jegerlehner (FJ)~\cite{Fleischer:1980ub} scheme. A detailed description of the scheme and its 
consequences for gauge independence can be found in~\cite{Krause:2016xku, Krause:2017mal,Altenkamp:2017ldc,Denner:2018opp}.
In the following sections we just present a brief description of the renormalization of the model sector by sector, giving the expressions required for the renormalization of the direct detection
process $\chi p \to \chi p$.

\subsection{Scalar Sector}\label{sec:ren-scalar-sector}

In the DSP of the N2HDM, after EWSB there are four neutral scalars (two CP-even, one CP-odd and the DM candidate $\chi$) and one charged Higgs pair. The OS conditions for the physical Higgs states result in the following mass counterterms,
\begin{align}
\label{eq:scalar-mass-CT}
 \delta m_{h_i}^2 =& \, \text{Re}[\Sigma_{h_i h_i}^\text{tad}(m_{h_i}^2)]\,, \, i \in \lbrace 1,2 \rbrace, \qquad \delta m_{\chi}^2 = \, \text{Re}[\Sigma_{\chi \chi}^\text{tad}(m_{\chi}^2)] \,,\nonumber  \\
\delta m_A^2 =& \, \text{Re}[\Sigma_{AA}^\text{tad}(m_{A}^2)]\,, \qquad \delta m_{H^\pm}^2 = \, \text{Re}[\Sigma_{H^\pm H^\pm}^\text{tad}(m_{H^\pm}^2)]\,,
\end{align}
where $\Sigma^{\text{tad}}(p^2)$ are the self-energies containing all tadpole topologies. Since the tadpoles are absorbed into the self-energies, explicit tadpole counterterms do not appear in the mass counterterms~\cite{Krause:2017mal}. 
The fields are renormalized in terms of the WFR constants $\delta Z_{\phi_i \phi_j}$ as 
\begin{equation}
\scriptstyle \begin{pmatrix}
\phi_i \\ \phi_j
\end{pmatrix}
\rightarrow
\begin{pmatrix}
1+\frac{1}{2}\delta Z_{\phi_i \phi_i} & \frac{1}{2}\delta Z_{\phi_i\phi_j}    \\
\frac{1}{2} \delta Z_{\phi_j \phi_i}  & 1+\frac{1}{2} \delta Z_{\phi_j\phi_j}
\end{pmatrix}
\begin{pmatrix}
\phi_i \\\phi_j
\end{pmatrix}\,,
\label{eq:renorm:ZfactorHiggs}
\end{equation}
where $\lbrace\phi_i, \phi_j\rbrace = \lbrace h_1, h_2\rbrace, \lbrace G, A\rbrace$ or $\lbrace G^\pm, H^\pm\rbrace$. We will just show explicitly the 2$\times$2 WFR matrices $\delta Z_{\phi_i, \phi_j}$ for
the case $\lbrace h_i, h_j\rbrace$ which reads
\beq
\label{eq:CPeven-WFR}
\scriptstyle
\left( \begin{array}{cc} \delta Z_{h_1 h_1} & \delta Z_{h_1 h_2} \\
	\delta Z_{h_2 h_1} & \delta Z_{h_1 h_1} \end{array} \right)
\hspace*{-0.2cm} &=& \hspace*{-0.2cm} 
\scriptstyle \left( \begin{array}{ccc} - \mbox{Re} \left.\frac{\partial
		\Sigma^{\text{tad}}_{h_1h_1} (p^2)}{\partial p^2}\right|_{p^2=m_{h_1}^2} &
	2 \frac{\mbox{Re} \left[\Sigma^{\text{tad}}_{h_1h_2} (m_{h_2}^2)\right]}{m_{h_1}^2-m_{h_2}^2} 
	   
	\notag \\[0.3cm] 
	2 \frac{\mbox{Re} \left[\Sigma^{\text{tad}}_{h_2h_1} (m_{h_1}^2)\right]}{m_{h_2}^2-m_{h_1}^2} & 
	- \mbox{Re} \left.\frac{\partial \Sigma^{\text{tad}}_{h_2h_2} (p^2)}{\partial p^2}\right|_{p^2=m_{h_2}^2} \\ 
\end{array} \right)\,,
\eeq
and the other two cases are obtained by replacing $\lbrace h_1, h_2\rbrace$ by $\lbrace G, A\rbrace$ and $\lbrace G^\pm, H^\pm\rbrace$.

%\beq
%\left( \begin{array}{cc} \delta Z_{G^0 G^0} & \delta Z_{G^0 A} \\
%	\delta Z_{A G^0} & \delta Z_{AA} \end{array} \right)
%\hspace*{-0.2cm} &=& \hspace*{-0.2cm} 
%\left( \begin{array}{cc} - \mbox{Re} \left.\frac{\partial
%		\Sigma^{\text{tad}}_{G^0 G^0} (k^2)}{\partial
%		p^2}\right|_{p^2=0} & -2 \frac{\mbox{Re}
%		\left[\Sigma^{\text{tad}}_{G^0 A} (m_A^2)\right] }{m_A^2}
%	\\[0.3cm] 
%	2 \frac{\mbox{Re} \left[ \Sigma^{\text{tad}}_{G^0 A} (0) \right]}{m_A^2} & 
%	- \mbox{Re} \left.\frac{\partial \Sigma^{\text{tad}}_{AA}
%		(p^2)}{\partial p^2}\right|_{p^2=m_A^2} 
%\end{array} \right)\,, \label{eq:CPodd-WFR} \\[0.2cm]
%\left( \begin{array}{cc} \delta Z_{G^\pm G^\pm} & \delta Z_{G^\pm H^\pm} \\
%	\delta Z_{H^\pm G^\pm} & \delta Z_{H^\pm H^\pm} \end{array} \right) 
%\hspace*{-0.2cm} &=& \hspace*{-0.2cm}
%\left( \begin{array}{cc} - \mbox{Re} \left.\frac{\partial
%		\Sigma^{\text{tad}}_{G^\pm G^\pm} (k^2)}{\partial
%		p^2}\right|_{p^2=0} & -2 \frac{\mbox{Re}
%		\left[\Sigma^{\text{tad}}_{G^\pm H^\pm} (m_{H^\pm}^2)
%		\right]}{m_{H^\pm}^2} \\[0.3cm] 
%	2 \frac{\mbox{Re} \left[ \Sigma^{\text{tad}}_{G^\pm H^\pm} (0)
%		\right]}{m_{H^\pm}^2} & 
%	- \mbox{Re} \left.\frac{\partial \Sigma^{\text{tad}}_{H^\pm H^\pm} (p^2)}{\partial
%		p^2}\right|_{p^2=m_{H^\pm}^2} 
%\end{array} \right)\,. 
%\label{eq:charged-WFR}
%\eeq
Finally the field strength renormalization for the DM particle $\chi$ is expressed in terms of its self-energy $\Sigma_{\chi\chi} (p^2)$ as
\be
\delta Z_{\chi \chi} =  - \mbox{Re} \left.\frac{\partial
	\Sigma^{\text{tad}}_{\chi\chi} (p^2)}{\partial p^2}\right|_{p^2=m_{\chi}^2}\,.
\label{eq:DM-WFR}
\ee

\subsection{Gauge Sector}\label{sec:ren-gauge-sector}
The gauge sector of the model is renormalized through OS conditions. The masses,  couplings and fields are expressed in terms of their counterterms as
\begin{align}
&m_W^2 \rightarrow \, m_W^2 + \delta m_W^2\,, \quad m_Z^2 \rightarrow \, m_Z^2 + \delta m_Z^2\,,\\
&e \rightarrow \, e + \delta Z_e\,, \quad g \rightarrow \, g + \delta g\,,\\
&W^{\pm}\to \, \cbrak{1 + \frac{1}{2} \delta Z_{WW}} W^{\pm}\,,\\
&\begin{pmatrix}
Z \\\gamma
\end{pmatrix}
\rightarrow\,
\begin{pmatrix}
1+\frac{1}{2} \delta Z_{ZZ}    & \frac{1}{2} \delta Z_{Z\gamma}          \\
\frac{1}{2}\delta Z_{\gamma Z} & 1 + \frac{1}{2} \delta Z_{\gamma\gamma}
\end{pmatrix}
\begin{pmatrix}
Z \\ \gamma
\end{pmatrix}\,,
\end{align}
where  $m_W$ and $m_Z$ are the $W$ and $Z$ boson masses, respectively, $e$ is the electric charge and $g$ is the weak $SU(2)_L$ coupling. The OS conditions for the masses give rise to the counterterms
\begin{align}
\delta m_W^2 = \mbox{Re} \Sigma^{\text{tad},T}_{WW} (m_W^2) \quad
\mbox{and} \quad
\delta m_Z^2 = \mbox{Re} \Sigma^{\text{tad},T}_{ZZ} (m_Z^2) \;, 
\end{align}
where the superscript $T$ indicates the transverse part of the self-energy, which also includes tadpole contributions. The counterterm for the electric charge is fixed in the Thomson limit
as in the SM and is expressed in terms of the Weinberg angle $\theta_W$ as
\begin{align}
\delta Z_e = \frac{1}{2} \left.\frac{\partial
	\Sigma^T_{\gamma\gamma} (k^2)}{\partial p^2}\right|_{p^2=0} +
\frac{\sin \theta_W}{\cos \theta_W} \frac{\Sigma_{\gamma Z}^T (0)}{m_Z^2} \,.
\label{eq:deltaze}
\end{align}
Using the above expression we can then fix the counterterm $\delta g$ as 
\be
 \frac{\delta g}{g} = \delta Z_e + \frac{1}{2} \frac{1}{m_Z^2-m_W^2} \cbrak{\delta m_W^2 -  \delta m_Z^2 \cos^2 \theta_W }\,.
 \ee
Finally, the WFR constants for the gauge fields are given by
\beq
\delta Z_{WW} &=& - \mbox{Re} \left.\frac{\partial \Sigma^{\text{tad},T}_{WW}
	(p^2)}{\partial p^2}\right|_{p^2=m_W^2}\,, \label{eq:sigmaww} \\
\left( \begin{array}{cc} \delta Z_{ZZ} & \delta Z_{Z\gamma} \\ \delta Z_{\gamma Z} &
	\delta Z_{\gamma\gamma} \end{array} \right) &=&
\left( \begin{array}{cc} - \mbox{Re} \left.\frac{\partial \Sigma^{\text{tad},T}_{ZZ}
		(p^2)}{\partial p^2}\right|_{p^2=m_Z^2} & 
	2 \frac{\Sigma^{\text{tad},T}_{Z\gamma} (0)}{m_Z^2} \\
	-2 \mbox{Re} \frac{\Sigma^{\text{tad},T}_{Z\gamma} (m_Z^2)}{m_Z^2} & 
	- \left.\frac{\partial \Sigma^{\text{tad},T}_{\gamma\gamma} (p^2)}{\partial
		p^2}\right|_{p^2=0} \end{array}\right) \;. \label{eq:sigmagamz}
\eeq

\subsection{Quark Sector}\label{sec:ren-quark-sector}
In the quark sector the OS scheme is applied for each quark. The renormalized quark fields are expressed in terms of their left- and right-handed components, with counterterms for each component as follows
\begin{eqnarray}
q_{L/R} \rightarrow \cbrak{1 + \frac{1}{2} \delta Z^{L/R}_{qq}} q \,,
\end{eqnarray}
with $q \in \lbrace u,d,s,c, b,t\rbrace$. In order to fix the counterterms we need to define the structure of the quark self-energies,
\be
\Sigma_q (p^2) = \slash{\!\!\! p} \Sigma^L_{qq} (p^2) P_L + \slash{\!\!\!
	p} \Sigma^R_{qq}
(p^2) P_R + m_f (P_L + P_R) \Sigma^{S}_{qq} (p^2)\,, 
\label{eq:quark-self}
\ee
where the self-energy superscripts $L, R$ and $S$ respectively
correspond to the left-handed, right-handed and scalar parts of the
quark self-energies, and $P_{L,R}$ are the left- and right-handed
projectors. Using the above expression the quark WFR constants and
mass counterterms in terms of the self-energies
  containing the tadpole topologies, are defined as
\begin{align}
\delta Z^{L/R}_q =& \, -\mbox{Re} \Sigma^{\text{tad},L/R}_q (m_q^2) 
- m_q^2
\frac{\partial}{\partial p^2} \mbox{Re} \left.\left(
\Sigma^{\text{tad},L/R}_{qq} (p^2) + 
\Sigma^{\text{tad},R/L}_{qq} (p^2) + 2\Sigma^{\text{tad},S}_{qq} (p^2) 
\right)\right|_{p^2=m_q^2} \,, \nonumber\\
\frac{\delta m_q}{m_q} =& \, \frac{1}{2} \mbox{Re} \left[ \Sigma^{\text{tad},L}_{qq}
(m_q^2) + \Sigma^{\text{tad},R}_{qq} (m_q^2) + 2\Sigma^{\text{tad},S}_{qq} \right] \,.
\end{align}

\subsection{Renormalization of the Mixing Angles}\label{sec:ren-mixing-angle}

Following the renormalization prescription for mixing angles in the 2HDM, the angles $\alpha$ and $\beta$ are renormalized as proposed in~\cite{Pilaftsis:1997dr, Kanemura:2004mg, Krause:2017mal}. The scheme connects
$\delta \alpha$ and $\delta \beta$ to the off-diagonal WFR constants of the scalar sector. Following again~\cite{Krause:2017mal} the angle counterterms are   

\begin{eqnarray}
\delta \alpha  =  \frac{1}{4}\cbrak{\delta Z_{h_1h_2}-\delta Z_{h_2h_1}}\,.
\end{eqnarray}
The counterterm for $\delta \beta$ can be derived either from the charged sector or the CP-odd sector using the same steps, and therefore we have two possible expressions for $\delta \beta$ given by
\beq
\delta \beta^{(1)} = \frac{1}{4} (\delta Z_{G^\pm H^\pm} -  \delta
Z_{H^\pm G^\pm}) \label{eq:betact1}
\eeq
and
\beq
\delta \beta^{(2)} &=& \frac{1}{4} (\delta Z_{G^0 A} -  \delta Z_{A
	G^0}) \label{eq:betact2} \;.
\eeq

\subsection{Renormalization of $\lambda_{7}$ and $\lambda_{8}$}\label{sec:ren-lam78}

We are left with the parameters $\lambda_{7}$ and $\lambda_{8}$ to complete the renormalization of the model. We will use a process-dependent scheme with the on-shell decays $h_i \to \chi \chi$ ($i=1,2$). 
The NLO amplitude $\mathcal{A}^{NLO}$ consists of the LO decay
amplitude $\mathcal{A}^{LO}$, the vertex corrections
$\mathcal{A}^{VC}$ and the counterterm amplitude $\mathcal{A}^{CT}$,
\beq
\mathcal{A}_{h_i}^{NLO}=\mathcal{A}_{h_i}^{LO}+\mathcal{A}_{h_i}^{VC}+\mathcal{A}_{h_i}^{CT},
\eeq
where the index $h_i$ denotes the decaying particle. 
The renormalization condition is such that we force the LO decay width
to be equal to NLO decay width.  
With the Higgs coupling $C_{\chi \chi h_i}$ between $h_i$ and two DM
particles given by
\beq
C_{\chi \chi h_i} = \left\{ \begin{array}{ll}
(\lambda_7 \cos\beta \cos\alpha + \lambda_8 \sin\beta \sin\alpha)
                              \frac{2m_W}{g}\,, & i=1 \\
(-\lambda_7 \cos\beta \sin\alpha + \lambda_8 \sin\beta \cos\alpha) 
                              \frac{2m_W}{g}\,, & i=2
\end{array} \right.
\label{eq:trilcoup}
\eeq
this condition gives rise to a system of equations for $\delta
\lambda_7$ and $\delta \lambda_8$ such that 
\begin{eqnarray}
 \frac{\partial \mathcal{A}_{h_1}^{LO}}{\partial \lambda_7} \delta \lambda_7+\frac{\partial \mathcal{A}_{h_1}^{LO}}{\partial \lambda_8}\delta \lambda_8  & = & 
  -\left( \mathcal{A}_{h_1}^{VC}+
                                                                                                                                                               \frac{1}{2}(\mathcal{A}_{h_2}^{LO}\delta Z_{h_2 h_1}+ \mathcal{A}_{h_1}^{LO}\delta Z_{h_1 h_1}+ 2 \mathcal{A}_{h_1}^{LO} \delta Z_{\chi \chi}) \right. \nonumber \\
& & \left. + \frac{ \partial \mathcal{A}_{h_1}^{LO}}{\partial m_W^2} \delta m_W^2 +\frac{ \partial \mathcal{A}_{h_1}^{LO}}{\partial g} \delta g + \frac{ \partial \mathcal{A}_{h_1}^{LO}}{\partial \alpha} \delta \alpha 
+ \frac{ \partial \mathcal{A}_{h_1}^{LO}}{\partial \beta} \delta \beta \right) \\
\frac{\partial \mathcal{A}_{h2}^{LO}}{\partial \lambda_7} \delta \lambda_7+\frac{\partial \mathcal{A}_{h2}^{LO}}{\partial \lambda_8}\delta \lambda_8  & = & 
-\left( \mathcal{A}_{h_2}^{VC}+ \frac{1}{2}(\mathcal{A}_{h_1}^{LO}\delta Z_{h_1 h_2}+ \mathcal{A}_{h_2}^{LO}\delta Z_{h_2 h_2}+ 2\mathcal{A}_{h_2}^{LO} \delta Z_{\chi \chi})  \right.\nonumber \\
& &+ \left. \frac{ \partial \mathcal{A}_{h_2}^{LO}}{\partial m_W^2} \delta m_W^2 +\frac{ \partial \mathcal{A}_{h_2}^{LO}}{\partial g} \delta g + \frac{ \partial \mathcal{A}_{h_2}^{LO}}{\partial \alpha} \delta \alpha + \frac{ \partial \mathcal{A}_{h_2}^{LO}}{\partial \beta} \delta \beta \right) \, ,
\end{eqnarray}
and this concludes our renormalization programme. We can now proceed to the calculation of the EW corrections.

\section{Electroweak Corrections to the SI Cross Section}\label{sec:cxsi}

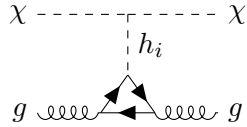
\begin{figure}[h]
    \centering
    \begin{tikzpicture}[baseline={([yshift=-.6ex]virtual.base)}]
        \definecolor{blobColor}{RGB}{191,191,191}
        \begin{feynman}
            \vertex (v1);
            \vertex[below=0.5cm of v1] (virtual);
            \vertex[below=0.4cm of v1] (blob);
            \vertex[below=0.8cm of virtual] (v2);
            \vertex[above=0.5cm of v2] (triangleTip);
            \vertex[left=0.36 of v2] (triangleLeft);
            \vertex[right=0.36cm of v2] (triangleRight);
            \vertex[left=1.2cm of v1] (chii) {\(\chi\)};
            \vertex[right=1.2cm of v1] (chio) {\(\chi\)};
            \vertex[left=1.2cm of v2] (qi) {\(g\)};
            \vertex[right=1.2cm of v2] (qo) {\(g\)};
            \diagram*{
            (chii) -- [scalar] (v1) -- [scalar] (chio),
            (v1) -- [scalar, edge label=\(h_i\)] (triangleTip),
            (qi) -- [gluon] (triangleLeft),
            (triangleRight) -- [fermion] (triangleLeft),
            (triangleRight) -- [gluon] (qo),
            (triangleLeft) -- [fermion] (triangleTip) -- [fermion] (triangleRight)
            };
        \end{feynman}
    \end{tikzpicture}
    \caption{Interaction of a DM particle and a gluon via a Higgs
      boson mediator and a quark loop.}
    \label{fig::gluonViaQuarkTriangle}
\end{figure}

The spin-independent (SI) DM-nucleon cross section can be written in terms of an effective 
coupling, $\alpha_n$, such that 
\begin{equation}
    \begin{tikzpicture}[baseline={([yshift=-.6ex]v.base)}]
        \begin{feynman}
            \vertex[circle, draw=black, fill=blobColor, minimum size=1.0cm] (v) {\(\alpha_n\)};
            \vertex[left=1.2cm of v] (virtual1);
            \vertex[right=1.2cm of v] (virtual2);
            \vertex[above=0.5cm of virtual1] (chii) {\(\chi\)};
            \vertex[above=0.5cm of virtual2] (chio) {\(\chi\)};
            \vertex[below=0.5cm of virtual1] (qi) {\(n\)};
            \vertex[below=0.5cm of virtual2] (qo) {\(n\)};
            \diagram*{
            (chii) -- [scalar] (v) -- [scalar] (chio),
            (qi) -- (v),
            (v) -- (qo),
            };
        \end{feynman}
    \end{tikzpicture}
    = i\mathcal{A}_n = i\alpha_n \overline{u}_nu_n = i\cdot 2m_n\alpha_n\,,
    \label{eq::effDMnuk}
\end{equation}
 where $ \overline{u}_nu_n = 2m_n$ ($m_n$ is the nucleon mass) because we assume that the velocity of the DM particle is negligibly small. With this definition the DM-nucleon cross section takes the form
\begin{equation}
    \sigma_n = \frac{1}{4\pi} \cbrak{\frac{m_n}{m_n+m_{\X}}}^2 \left|\alpha_n\right|^2\,,
\end{equation}
where $\mX$ is the DM
mass. As the nucleon is a bound state, the DM-nucleon coupling receives contributions
both from valence quarks  $\cbrak{q=u,d,s}$ and from the gluons.  
The SI DM-nucleon cross section is calculated using a parton basis,
with the operators considered
in the non-relativistic limit. Its most general form is given by~\cite{Hisano:2015bma}
\begin{equation}
    \mathcal{L}_{\text{eff}} = \sum_q C_S^q\mathcal{O}^q_S + C^g_S \mathcal{O}^g_S + \sum_q C_T^q \mathcal{O}^q_T\,,
    \label{eq:LeffParton}
\end{equation}
with the operators
\begin{subequations}
    \begin{eqnarray}
        \mathcal{O}_S^q = m_q \X^2 \bar{q}q\,,\\
        \mathcal{O}^g_S = \frac{\alpha_s}{\pi} \X^2 \GG\,,\\
        \mathcal{O}^q_T = \frac{1}{\mX^2} \X^2 \ii \partial^{\mu} \ii \partial^{\nu} \underbrace{\frac{1}{2}\ii\bar{q}\cbrak{\partial_{\mu}\gamma_{\nu}+\partial_{\nu}\gamma_{\nu}-\frac{1}{2}g_{\mu\nu}\slashed{\partial}}q}_{\equiv \mathcal{O}_{\mu\nu}^{q}}\, ,
    \end{eqnarray}
    \label{eq::partonoperators}
\end{subequations}
\noindent which are built with the DM field $\X$, the quark spinor $q$ and the gluon field strength tensor
$G^a_{\mu\nu}$; $\alpha_s$ is the strong coupling constant. The quark-DM interaction is encoded in the operator
$\mathcal{O}^q_S$ while the gluon-DM interaction is encoded in  $\mathcal{O}^g_S$. Finally, the twist-2 operator $\mathcal{O}_{\mu\nu}^{q}$
also contributes to the SI cross section.
The expectation values of the operators in \cref{eq::partonoperators} are written as~\cite{Hisano:2012wm,Young:2009zb,Shifman:1978zn}
\begin{subequations}
    \begin{eqnarray}
        \bra{n}m_q\bar{q}q\ket{n} &\equiv& m_n f^n_q \,,\\
        \bra{n}-\frac{\alpha_s}{12\pi}\GG\ket{n} &\equiv& \frac{2}{27}m_n f_g^n \,,
    \end{eqnarray}
    \label{eq::nucmatrix}
\end{subequations} 
\noindent where the nucleon matrix elements $f^n_q$ and $f^n_g$ 
are determined by lattice calculations. Their numerical values are
given in \cref{sec::NumericalSetUp}. The QCD trace anomaly relates the heavy
quark $Q=b,c,t$ operators with the gluon field strength tensor~\cite{Shifman:1978zn}
\begin{equation}
    m_Q \bar{Q}Q \rightarrow -\frac{\alpha_s}{12 \pi}\GG\,,
    \label{eq::Mapping}
\end{equation}
corresponding to the Feynman diagram in~\cref{fig::gluonViaQuarkTriangle} and can therefore be determined by first
calculating the  (heavy) external quark process and then using
\cref{eq::Mapping} to determine the effective gluon interaction. These
amplitudes are then used to the extract the Wilson coefficients
$C^g_S$.

The DM nucleon cross section $\sigma_n$ is given at NLO by
\beq
\sigma_n=\frac{1}{\pi} \left( \frac{m_n}{m_{\chi}+m_n} \right)^2 \left( |f_n^{LO}|^2+2 Re\left( f_n^{LO} f_n^{NLO}\right) \right)
\eeq
where, according to the previous discussion, the LO and NLO form factors are given as
\beq
f_n^{LO}= m_n C_S^q \left( \sum_{q=u,d,s} f_{q}^n + \sum_{q=c,b,t}\frac{2}{27} f_{g}^{n} \right)\, ,
\eeq
\beq
f_n^{NLO}= m_n  \left( \sum_{q=u,d,s} C_S^{q,NLO} f_{q}^n
  +\sum_{q=u,d,s,c,b} \frac{3}{4} \left( q(2)+\bar{q}(2) \right)C^q_T-
  \frac{8 \pi}{9 \alpha_s} C_S^g f_{g}^n  \right) \, .
\eeq
We will neglect the term proportional to $C_S^g$  because the matching in~\cref{eq::Mapping} cannot be used at EW NLO. As discussed in~\cite{Glaus:2020ihj}, taking into account 
the EW NLO corrections of the gluon contributions would require a proper mixed QCD-EW matching of the QCD trace anomaly. Therefore the NLO contributions considered are 
\beq
C_S^{q,NLO} = f_q^{uV}+f_q^{lV}+f_q^{med}+f_q^{box} \, ,\\
C_T^q = g_q^{box} \;.
\eeq
Here $f_q^{i}$ are the Wilson coefficients from the upper vertex corrections, lower vertex corrections, mediator corrections and the box contributions. The $g_q^{box}$
 are the box contributions proportional to the second momenta of the
 quarks $q^n(2)$. The values for $q^n(2)$ are also given in
 \cref{sec::NumericalSetUp}. 

\subsection{Upper vertex corrections}\label{sec:cxsiuV}

For the extraction of the Wilson coefficients of the upper vertex the
one-loop corrections to the coupling $\chi \chi h_i$ ($i=1,2$) need to
be calculated. For this purpose $\chi$ is taken on-shell and it is
assumed that the momentum transfer goes to zero. Thus the incoming
momentum $p_{in}$ of the dark matter particle equals the outgoing
momentum $p_{out} \equiv p$. The NLO amplitude
consists of the LO amplitude 
$\mathcal{A}^{LO}$, the virtual vertex corrections $\mathcal{A}^{VC}$
and the counterterm $\mathcal{A}^{CT}$. In the limit of zero momentum
transfer the LO amplitudes for the upper vertex
topology read ($i=1,2$)
%
%\beq 
%i\mathcal{A}^{LO}_{h_1} &= -\lambda_7 \cos \alpha\, \cos \beta- v \lambda_8 \sin \alpha \, \sin \beta \, , \\
%i\mathcal{A}^{LO}_{h_2} &= -\lambda_7 \sin \alpha \,  \cos \beta- v \lambda_8 \cos \alpha \, \sin \beta. 
%\eeq 
%
\beq 
i\mathcal{A}^{LO}_{h_i} &=& C_{\chi \chi h_i} C_{qqh_i}
\frac{1}{m_{h_i}^2} \bar{u}(p) u(p)\,, 
\eeq 
with $C_{\chi \chi h_i}$ given in Eq.~(\ref{eq:trilcoup}), the Higgs
$h_i$ coupling to a quark pair given by
\beq
C_{qqh_i} = - \frac{gm_q}{2m_W} \frac{R_{\alpha,i2}}{\sin\beta}
\eeq
and $u(\bar{u})$ denoting the spinor of the (anti-)quark. 
The counterterm amplitudes read
%\beq
%i\mathcal{A}^{CT}_{h_1} =  \left( \frac{1}{2}(\mathcal{A}^{LO}_{h_2} \delta Z_{h_2 h_1}+\mathcal{A}^{LO}_{h_1}\delta Z_{h_1 h_1})+\mathcal{A}^{LO}_{h_1} \delta Z_{\chi \chi} + \delta C_{\chi\chi h_1} \right) \epsilon (p) \epsilon^* (p)\\
%i\mathcal{A}^{CT}_{h_2} =  \left( \frac{1}{2}(\mathcal{A}^{LO}_{h_1} \delta Z_{h_1 h_2}+\mathcal{A}^{LO}_{h_2}\delta Z_{h_2 h_2})+\mathcal{A}^{LO}_{h_2} \delta Z_{\chi \chi} + \delta C_{\chi\chi h_2} \right) \epsilon (p) \epsilon^* (p)
%\eeq
%
\beq
i\mathcal{A}^{CT}_{h_1} =  \frac{\mathcal{C}_{qq h_1}}{m_{h_1}^2} \left( \frac{1}{2}(C_{\chi\chi h_2} \delta Z_{h_2 h_1}+C_{\chi \chi h_1}\delta Z_{h_1 h_1})+C_{\chi\chi h_1} \delta Z_{\chi \chi} + \delta C_{\chi\chi h_1} \right)\\
i\mathcal{A}^{CT}_{h_2} = \frac{C_{qq h_2}}{m_{h_2}^2} \left( \frac{1}{2}(C_{\chi \chi h_1} \delta Z_{h_1 h_2}+C_{\chi \chi h_2}\delta Z_{h_2 h_2})+C_{\chi \chi h_2} \delta Z_{\chi \chi} + \delta C_{\chi\chi h_2} \right) \, ,
\eeq
where the counterterm of the coupling is obtained by varying the trilinear coupling
  $C_{\chi\chi h_i}$, 
%\beq
%\delta C_{\chi\chi h_i} = \frac{\partial \mathcal{A}^{LO}_{h_i}}{\partial m_{W}^2} \delta m_{W}^2+\frac{\partial \mathcal{A}^{LO}_{h_i}}{\partial \alpha} \delta \alpha+\frac{\partial \mathcal{A}^{LO}_{h_i}}{\partial \beta} \delta \beta+\frac{\partial \mathcal{A}^{LO}_{h_i}}{\partial g_2} \delta g_2+\frac{\partial \mathcal{A}^{LO}_{h_i}}{\partial \lambda_7} \delta \lambda_7+\frac{\partial \mathcal{A}^{LO}_{h_i}}{\partial \lambda_8} \delta \lambda_8.
%\eeq
\beq
\delta C_{\chi\chi h_i} = \frac{\partial C_{\chi\chi h_i}}{\partial
  m_{W}} \delta m_{W}+\frac{\partial C_{\chi\chi h_i}}{\partial
  \alpha} \delta \alpha+\frac{\partial C_{\chi\chi h_i}}{\partial
  \beta} \delta \beta+\frac{\partial 
  C_{\chi\chi h_i}}{\partial g_2} \delta g_2+\frac{\partial C_{\chi
    \chi h_i}}{\partial \lambda_7} \delta \lambda_7+\frac{\partial
  C_{\chi\chi h_i}}{\partial \lambda_8} \delta \lambda_8.
\eeq
Therefore the NLO amplitude at zero momentum transfer is given by 
\beq
i\mathcal{A}^{NLO}_{h_i} = i\mathcal{A}^{LO}_{h_i}  +  i\mathcal{A}^{VC}_{h_i}  + i\mathcal{A}^{CT}_{h_i} 
\eeq

 %\bar{u}(p_q) u(p_q)
\subsubsection{Mediator corrections}\label{sec:cxsiuV}
To obtain the Wilson coefficient from the mediator corrections the one-loop corrections to the propagator and the corresponding counterterms need to be determined. 
They can be expressed in terms of the renormalized one-loop propagator
$\cbrak{i,j=1,2}$ 
\begin{equation}
    \Delta_{h_ih_j} =
    -\frac{\hat{\Sigma}_{h_ih_j}(p^2=0)}{m_{h_i}^2m_{h_j}^2} \;,
\end{equation}
with the renormalised self-energy matrix
\begin{equation}
    \begin{pmatrix}
        \hat{\Sigma}_{h_1h_1} & \hat{\Sigma}_{h_1h_2} \\
        \hat{\Sigma}_{h_2h_1} & \hat{\Sigma}_{h_2h_2}
    \end{pmatrix}
    \equiv
    \hat\Sigma(p^2) = \Sigma(p^2) - \delta m^2 -\delta T +\frac{\delta Z}{2} \cbrak{p^2-\mathcal{M}^2} + \cbrak{p^2-\mathcal{M}^2}\frac{\delta Z}{2}\,,
\end{equation}
with $\delta T$ denoting the tadpole counterterm matrix and
$\mathcal{M}\equiv m_{h_i}^2 \delta_{ij}$ the diagonal mass matrix.
The details of the calculation can be found in~\cite{Glaus:2020ihj}. Here we just note that the self-energies receive extra contributions relative to the complex singlet extension in~\cite{Glaus:2020ihj} 
because there are new scalars in the loop from the second doublet.

\subsection{Lower vertex corrections }\label{sec:cxsilV}
The lower vertex corrections are also calculated exactly as in~\cite{Glaus:2020ihj}. The difference is again the contribution of the new scalar particles in the loop. However from the point of view 
of the renormalization procedure nothing changes. A very detailed discussion on the different problems arising in this calculation is presented in our Ref.~\cite{Glaus:2020ape}.  
Of particular importance is the treatment of infrared divergences and the discussion on the heavy quark contributions to the process. 

\subsubsection{Box diagrams}\label{sec:box}

Finally, the calculation of the box corrections is also discussed in
detail in our previous works, Refs.~\cite{Glaus:2019itb,
  Glaus:2020ape, Glaus:2020ihj} and again was shown to be one order of
magnitude smaller than the main contribution.

%%%%%%%%%%%%%%%%%%%%%%%%%%%%%%%%%%%%%%%%%
\section{Results and discussion}
\label{sec:num}

%%%%%%%%%%%% stop
One of the Higgs bosons, either $h_1$ or $h_2$, is the SM-like Higgs boson with a mass of $\SI{125.09}{GeV}$\cite{Aad:2015zhl}. 
The other CP-even Higgs can be lighter or heavier than the SM-like Higgs boson.
The points presented in the scatter plots were generated using \texttt{ScannerS}~\cite{Coimbra:2013qq, Muhlleitner:2020wwk} where the most relevant experimental and theoretical constraints were taken into account.  
\texttt{ScannerS} checks if the potential is bounded from below, that there is a global minimum and that perturbative unitarity holds.  Agreement with the electroweak precision measurements at the 2$\sigma$ level is enforced using 
the $S, T, U$ \cite{Peskin:1991sw, Grimus:2008nb}  parameters. Collider bounds from Tevatron, LEP and LHC are encoded in \texttt{HiggsBounds 5.6.0}~\cite{Bechtle:2020pkv} and
\texttt{HiggsSignals 2.3.1}~\cite{Bechtle:2013xfa}.  We ask for a 95\% confidence level agreement using the exclusion limits for all available searches for non-standard Higgs bosons, including Higgs invisible decays.
Branching ratios are calculated using \texttt{AnyHdecay 1.1.0}~\cite{Muhlleitner:2020wwk}. 
The code includes the Higgs decay widths for the N2HDM~{\tt
 N2HDECAY}~\cite{Engeln:2018mbg}, with state-of-the art higher-order
QCD corrections.  
  The code {\tt  N2HDECAY} is based on the implementation of the N2HDM in {\tt
  HDECAY}~\cite{Djouadi:1997yw,Djouadi:2018xqq}. All EW radiative corrections in  {\tt HDECAY} are turned off for consistency.

The DM relic abundance is calculated using \texttt{MicrOMEGAs}~\cite{Belanger:2013oya}, and a bound on its value is set
by the current experimental result $({\Omega}h^2)^{\rm obs}_{\rm DM} = 0.1186 \pm 0.002$ from the Planck
Collaboration~\cite{Ade:2015xua}. 
%We only exclude values for which the calculated relic density is above the observed
%central value plus 2$\sigma$. 
%
We require the calculated relic abundance to be equal or below its experimental central value plus 2$\sigma$, 
that is, we allow the DM  not to saturate the relic density and therefore define a DM fraction 
\begin{eqnarray}
f_{\chi\chi} = \frac{({\Omega} h^2)_{\chi}}{(\Omega h^2)^{\rm
  obs}_{\text{DM}}}\,, 
\label{eq:dmfraction}
\end{eqnarray} 
where $(\Omega h^2)_{\chi}$ stands for the calculated DM relic abundance in our model. 
As for direct detection, the XENON1T~\cite{Aprile:2017iyp,Aprile:2018dbl} experiment provides the most
stringent upper bound on the spin-independent DM nucleon scattering. 

\begin{figure}[h!]
\centering
\includegraphics[width = 0.44 \linewidth]{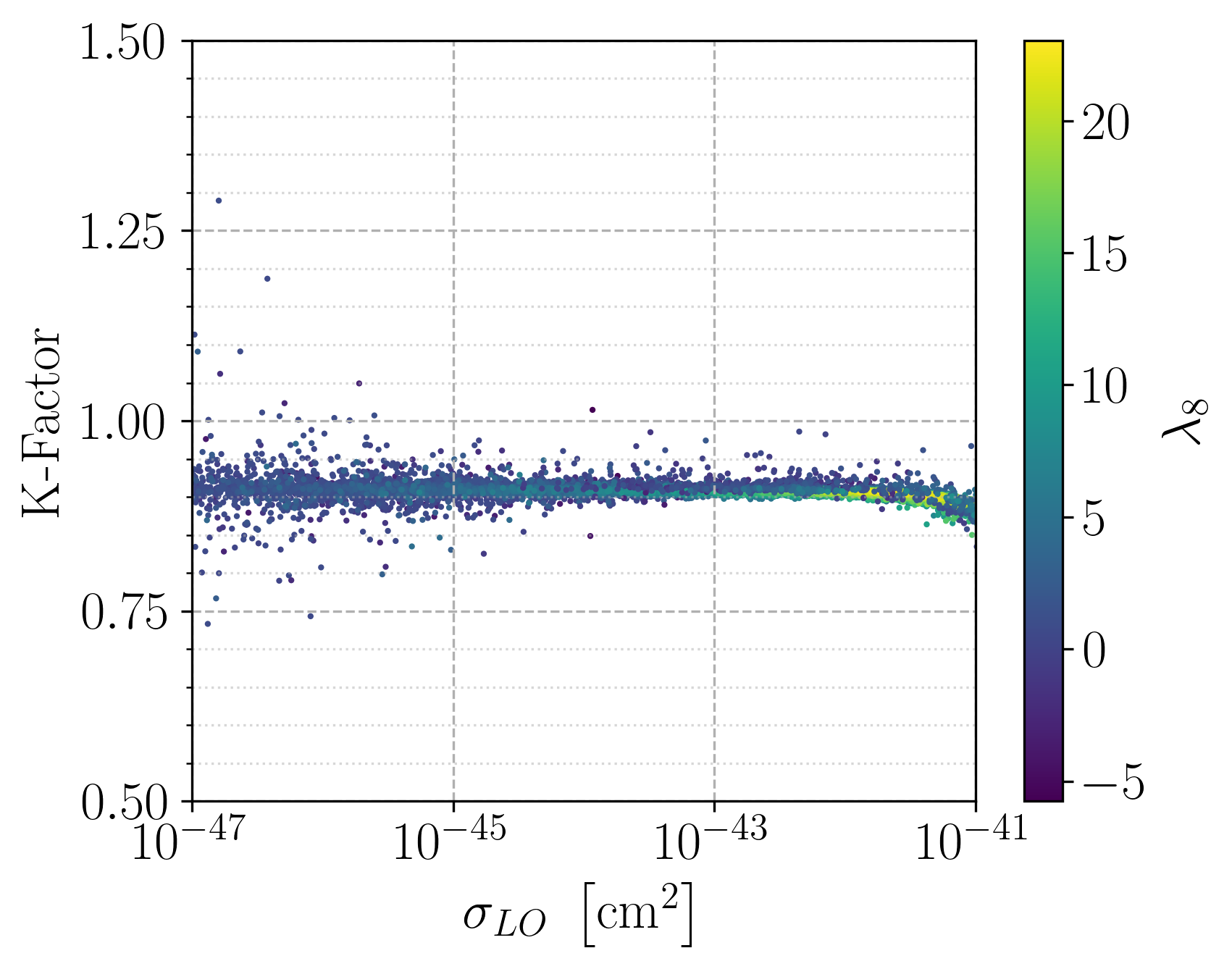}
\includegraphics[width = 0.44 \linewidth]{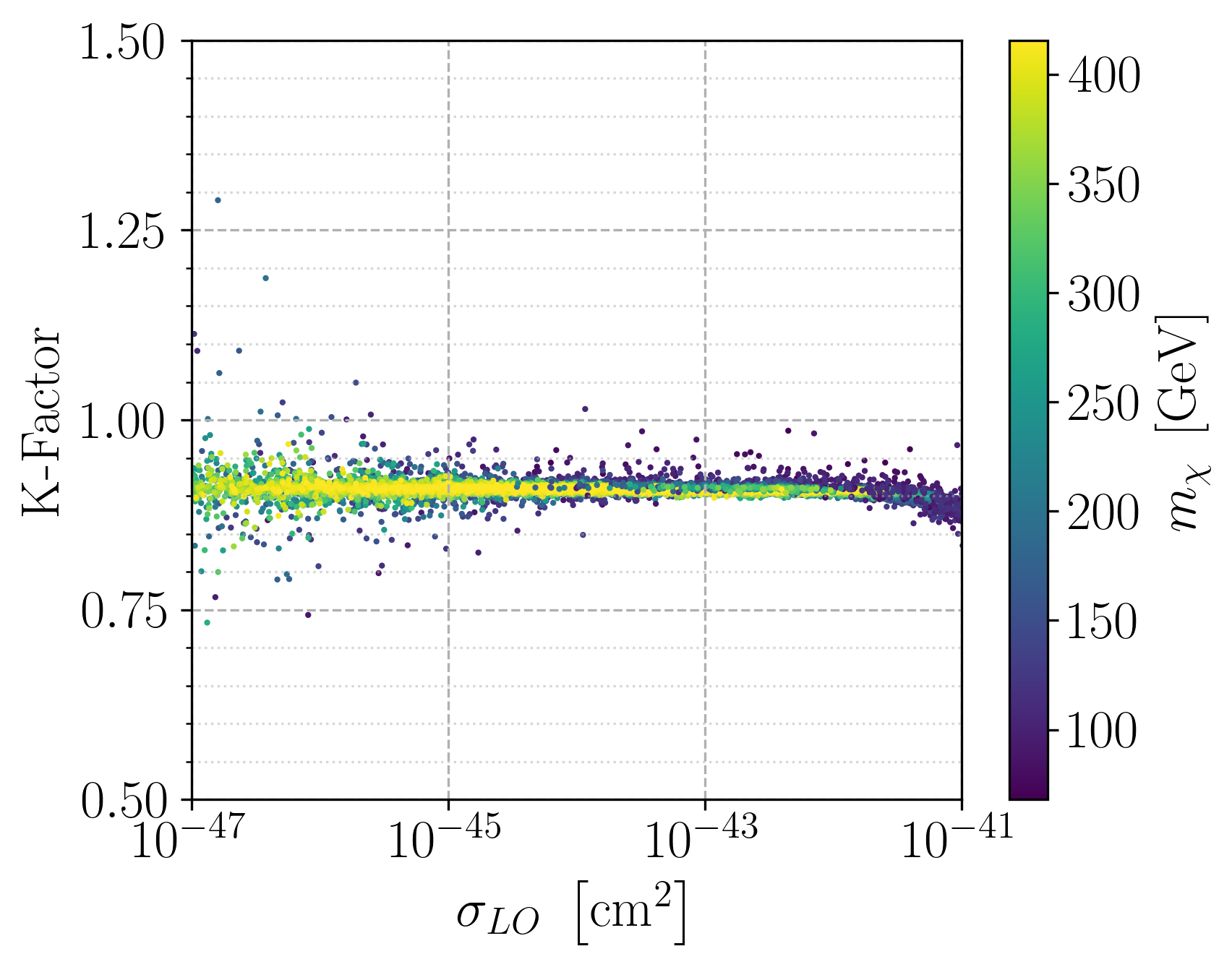}
\caption{Scatter plot showing the $K$-factor as a function of the $\sigma_{LO}$ DM-nucleon spin-independent cross section. In the left
panel we show the behaviour with $\lambda_8$ in the colour bar while
in the right panel the DM mass is now shown in the colour bar.}
\label{Figa}
\end{figure}

The ranges of the input parameters for the scan are shown in
Table~\ref{tab:vdmscan}\color{black}. $m_\Phi$ denotes the masses of $\Phi =
  h_i,A,H^\pm$, where $h_i$ is the non-125 GeV Higgs. Note that the
constraints are applied and therefore the allowed parameter space 
will be a small fraction of the initial space. 
\begin{table}[h!]
\begin{center}
\begin{tabular}{l|cccccc} \toprule
& $m_\Phi$ [GeV] & $\mX$ [GeV] & $\tan \beta$ & $\alpha$ & $\lambda_{7,8}$ \\ \hline
%& & & & \multicolumn{11}{c}{in GeV} \\ \midrule
min & 50 & 1 & 1 & $-\frac{\pi}{2}$  & $- 4 \pi$ \\
max & 1000 & 1000 & $30$ & $\frac{\pi}{2}$   & $4 \pi$\\ \bottomrule
\end{tabular}
\caption{Input parameters for the DSP of the N2HDM scan, all parameters
  varied independently between the given minimum and maximum
  values. The SM-like Higgs boson mass is set
  $m_h=125.09$~GeV and the SM VEV
  $v=246.22$~GeV. $m_\Phi$ denotes the masses of $\Phi =
  h_i,A,H^\pm$. The parameter $\lambda_6$ does not play any role for
  our computation.
  %\textcolor{blue}{Does the $10^7$ really apply to $v$   and not
  %rather to $v^2$?}\textcolor{orange}{indeed $v_S$! $\max v_S \sim 4\cdot10^6$ in the sample}
  \label{tab:vdmscan}}
\end{center}
\end{table}
%%%%%%%%%%%%%%%%%%%%%%%%%%%%%%%%%%%%%%%%%%%%%%%%%%%%%%%%%%%%%%%%%%%%%%%%%%%%%%%%%%%%%%%%%%%%
%
In Fig.~\ref{Figa} we present a scatter plot of the $K$-factor,
defined as $K=\sigma_{NLO/}\sigma_{LO}$, as a function of the
$\sigma_{LO}$ DM-nucleon spin-independent cross section. In the left 
panel we show the behaviour with $\lambda_8$ in the colour bar while
in the right panel the DM mass is shown in
the colour bar. The XENON1T experiment sets an upper
limit on the cross section of $10^{-45}$ cm${^2}$ valid for any value
of the DM mass (the limit is stronger for smaller DM
masses). Therefore, in the region of interest it is clear that the correction are very
stable with the bulk of the points just below $0.94$. The largest
value of the correction yields a $K = 1.12$ and the lowest 
value of $K$ is just below $0.75$. Hence the correction are stable and
there is in general a slight decrease of the cross section at NLO. We
have also checked that in the region of interest 
none of the free parameters play a special role in the $K$-factor values.
%
%%%%%%%%%%%%%%%%%%%%%%%%%%%%%%%%%%%%%%%%%%%%%%%%%%%%%%%
%%%%%%%%%%%%%%%%%%%%%%%%%%%%%%%%%%%%%%%%%%%%%%%%%%%%%%%

\begin{figure}[h!]
\centering
\includegraphics[width = 0.44 \linewidth]{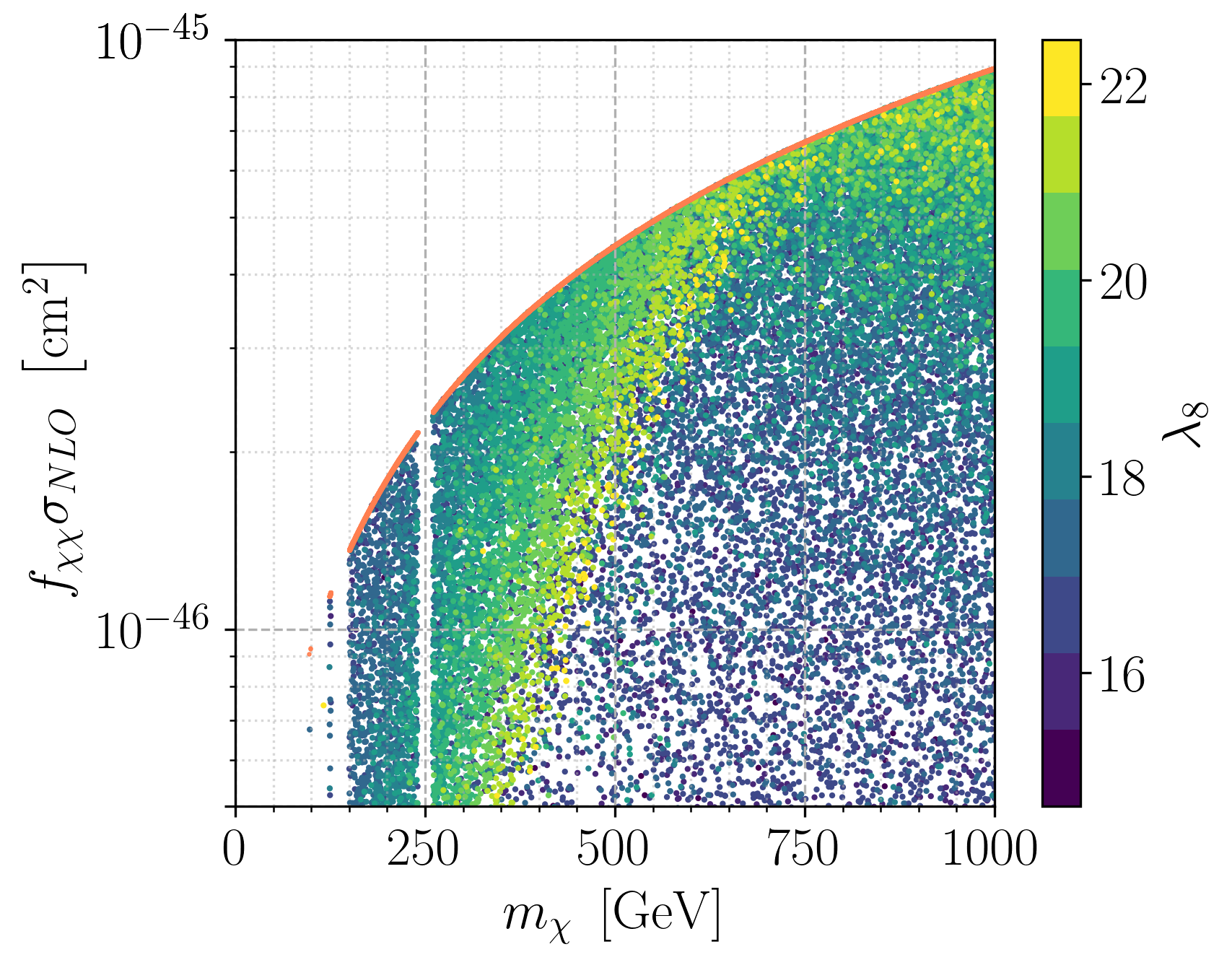}
\includegraphics[width = 0.44 \linewidth]{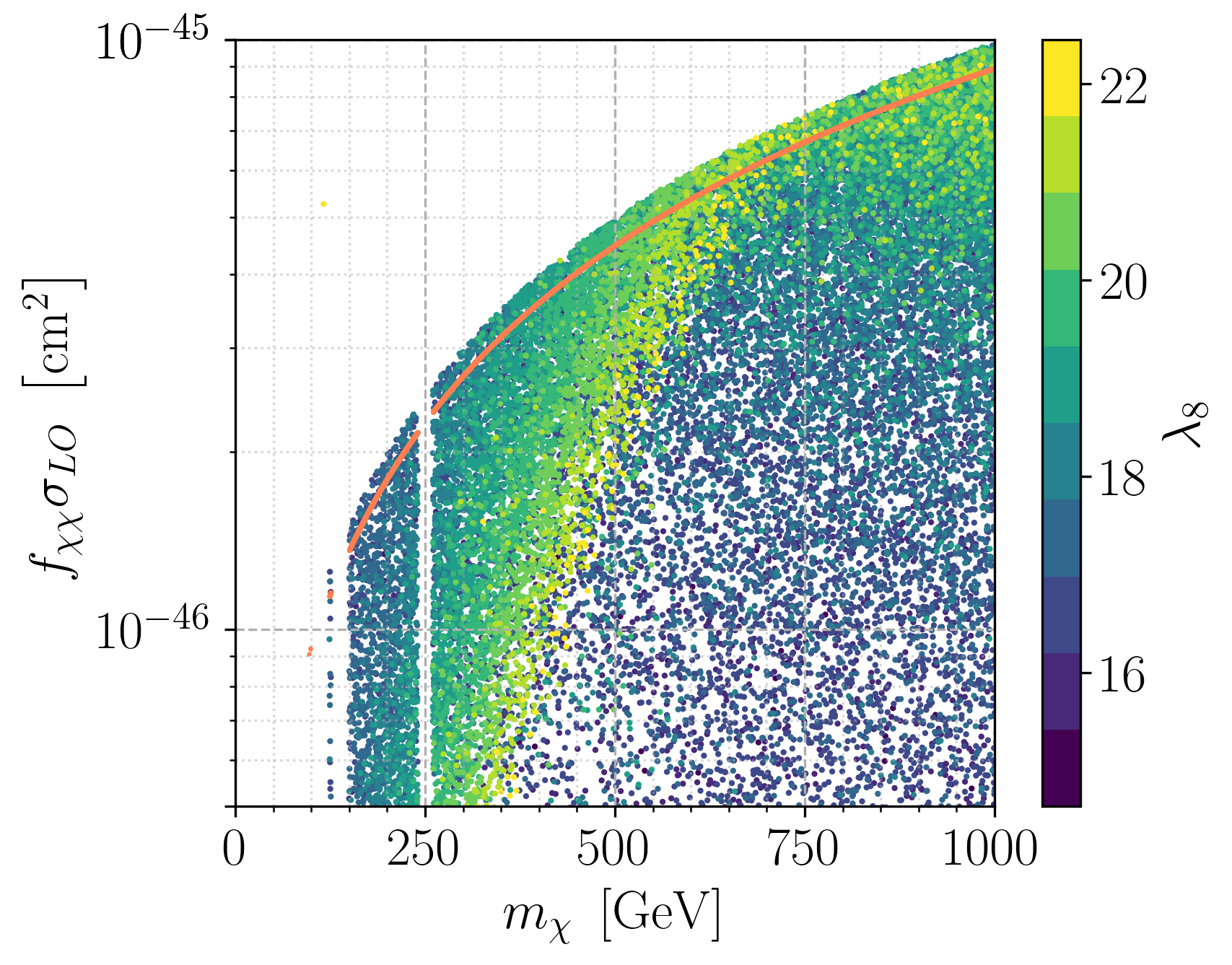}
\caption{Scatter plot showing the SI cross section, including the correction factor $f_{\chi\chi}$, at NLO (left) and LO (right) compared to the
      Xenon limit (orange line), as a function of the DM mass. The
      colour bar shows the dependence on the parameter $\lambda_8$.}
\label{Figb}
\end{figure}

In Fig.~\ref{Figb} we now show the DM-nucleon cross section including the correction factor $f_{\chi\chi}$, at NLO (left) and LO (right) compared to the XENON1T limit (orange line), as a function of the DM mass.
The points shown are such that they are all below the XENON1T line at
NLO as can be seen in the left plot. In the right plot we show that
some of the points would not comply with XENON1T limit if calculated
at LO. 
%%%%%%%%%%%%%%%%%%%%%%%%%%%%%%%%%%%%%%%%%%%%%%%%%%%%%%%
%%%%%%%%%%%%%%%%%%%%%%%%%%%%%%%%%%%%%%%%%%%%%%%%%%%%%%%

\section{Conclusions}
\label{sec:conc}

We have calculated  the spin-independent DM-nucleon scattering cross
section for the DSP of the N2HDM including higher-order corrections.
One of the main goals of this work was to check if the 
parameter space of the model compatible with the most important theoretical and
  experimental constraints would give rise to large and/or unstable
higher-order corrections. We found that in this model the corrections
are stable with a $K$-factor close to one. The reason behind this result is that the main corrections
come from the upper vertex, mediator and lower vertex, where the diagrams are similar to the ones in
the CxSM. In the CxSM, although the LO cross section turns out to be negligible due to a
peculiar Feynman diagram cancellation, the NLO corrections are quite stable
as discussed in ~\cite{Azevedo:2018exj, Ishiwata:2018sdi, Glaus:2020ihj}. 
The same is true for the vector dark matter (VDM) model discussed in~\cite{Glaus:2019itb, Glaus:2020ape}.
For the VDM the LO cross section is not negligible and the $K$-factor is quite stable and close to one,
except for large values of the dark gauge coupling.

Still, in the model discussed in this work, there are new particles in the loops, like for instance in the case of the CP-even self-energies which
contribute to the mediator correction, one could in principle expect
sizeable corrections which is not the case.
The masses and couplings in this model are already very constrained by the LHC results but a such a stable
result could not be anticipated.

From the phenomenological point of view the overall conclusions are the following.
The NLO corrections can increase the LO results to values where the XENON1T experiment becomes sensitive to
the model, or to values where the model is even excluded due to cross sections values above the XENON1T limit. 
But the reverse is also true even if not as common. Parameter points that might be rejected at LO may render 
the model viable when NLO corrections are included. We conclude that as a first approximation the LO
cross section is a very good approximation but if a DM candidate is detected NLO corrections should be taken
into account in order to either validate or exclude the model.

%%%%%%%%%%%%%%%%%%%%%%%%%%%%%%%%%%%%%%%%%%%%%%%%%%%%%%%
\bigskip
\bigskip
\subsubsection*{Acknowledgments}
RS is supported by FCT under contracts UIDB/00618/2020, UIDP/00618/2020, PTDC/FIS-PAR/31000/2017, CERN/FISPAR
/0002/2017, CERN/FIS-PAR/0014/2019. The work of MM is supported by the
BMBF-Project 05H21VKCCA.

\appendix

\section{Numerical Values for the paramaters}
\label{sec::NumericalSetUp}
In this appendix we present the numerical values of the parameters used in the calculation of the cross sections. The SM input parameters are~\cite{Dittmaier:2011ti}
\begin{equation}
    \begin{split}
        m_u &= \SI{0.19}{GeV}\,,		\qquad &m_c &= \SI{1.4}{GeV}\,,		\qquad &m_t &= \SI{172.5}{GeV}\,, \\
        m_d &= \SI{0.19}{GeV}\,,		\qquad &m_s &= \SI{0.19}{GeV}\,,		\qquad &m_b &= \SI{4.75}{GeV}\,, \\
        m_e &= \SI{0.511}{MeV}\,,		\qquad &m_\mu &= \SI{105.658}{MeV}\,,	\qquad &m_\tau &= \SI{1.777}{GeV}\,, \\
        m_W &= \SI{80.398}{GeV}\,, 	\qquad &v &= \SI{246}{GeV}\,, \\
        m_Z &= \SI{91.188}{GeV}\,. \\
    \end{split}
\end{equation}
The $SU(2)$ gauge coupling $g$ and the Weinberg angle are calculated as 
\begin{equation}
    g = 2m_W/v = \SI{0.653}{}\,, \qquad \sin \theta_W = m_W/m_Z = 0.472\,.
\end{equation}

The nucleon cross section is calculated for the proton, meaning $ \sigma  \equiv \sigma_p$, and the mass of the proton is $ m_p = \SI{0.938}{GeV}$.

The nuclear matrix elements for the proton have the following values~\cite{Hisano:2012wm,Young:2009zb}  
\begin{equation}
    \begin{split}
        f_u^p &= \SI{0.01513}{}\,,		\qquad &f_d^p &= \SI{0.0191}{}\,,		\qquad &f_s^p &= \SI{0.0447}{}\,, \\
        f_g^p &= \SI{0.92107}{}\,, \\
        u^p(2) &= \SI{0.22}{}\,,		\qquad &c^p(2) &= \SI{0.019}{}\,, \\
        \bar{u}^p(2) &= \SI{0.034}{}\,,		\qquad &\bar{c}^p(2) &= \SI{0.019}{}\,, \\
        d^p(2) &= \SI{0.11}{}\,,		\qquad &s^p(2) &= \SI{0.026}{}\,,	\qquad &b^p(2) &= \SI{0.012}{}\,, \\
        \bar{d}^p(2) &= \SI{0.036}{}\,,		\qquad &\bar{s}^p(2) &= \SI{0.026}{}\,,	\qquad &\bar{b}^p(2) &= \SI{0.012}{}\,,
    \end{split}
\end{equation}
and no uncertainties in the determination of these nuclear matrix elements were taken into account.

\bibliographystyle{JHEP}
\bibliography{dddm.bib}
\end{document}